# Control of fuel target implosion non-uniformity in heavy ion inertial fusion

T. Iinuma[1]*, T. Karino[1], S. Kondo[1], T. Kubo[1,] H. Kato[1], T. Suzuki[1], S. Kawata[1]*, A. I. Ogoyski[2]

[1] Utsunomiya University, Graduate School of Eng., Utsunomiya, 321-8585, Japan

[2] Varna Technical University, Dept. of Physics, Varna, 9010, Bulgaria

## Abstract

In inertial fusion, one of scientific issues is to reduce an implosion non-uniformity of a spherical fuel target. The implosion non-uniformity is caused by several factors, including the driver beam illumination non-uniformity, the Rayleigh-Taylor instability (RTI) growth, etc. In this paper we propose a new control method to reduce the implosion non-uniformity; the oscillating implosion acceleration $\delta g$(t) is created by pulsating and dephasing heavy ion beams (HIBs) in heavy ion inertial fusion (HIF). The $\delta g$(t) would reduce the RTI growth effectively. The original concept of the non-uniformity control in inertial fusion was proposed in (Kawata, *et al.*, 1993). In this paper it was found that the pulsating and dephasing HIBs illumination provide successfully the controlled $\delta g$(t) and that $\delta g$(t) induced by the pulsating HIBs reduces well the implosion non-uniformity. Consequently the pulsating HIBs improve a pellet gain remarkably in HIF.

*Corresponding author: mt156204@cc.utsunomiya-u.ac.jp, kwt@cc.utsunomiya-u.ac.jp

Introduction

A heavy ion beam (HIB) has preferable features to release the fusion energy in inertial fusion: in particle accelerators HIBs are generated with a high driver efficiency of ~ 30-40%, and the HIB ions deposit their energy inside of materials. Therefore, a requirement for the fusion target energy gain is relatively low, that would be ~50 to operate a HIF fusion reactor with the standard energy output of 1GW of electricity (Kawata, Karino & Ogoyski, 2016). Key issues in heavy ion inertial fusion (HIF) include a target implosion uniformity to obtain a sufficient fusion energy output. The requirement for the implosion uniformity is very stringent, and the implosion non- uniformity must be less than a few %(Kawata & Karino, 2015; Kawata *et al*, 2016; Emery *et al*, 1982; Kawata & Niu, 1984). Therefore, it is essentially important to improve the fuel target implosion uniformity. The target implosion should be robust against the implosion non- uniformities for the stable reactor operation. In general, the target implosion non-uniformity is introduced by the driver beams' illumination non-uniformity, an imperfect target sphericity, a non-uniform target density, a target alignment error in a fusion reactor, et al. To reduce the non-uniformity, we have focused on the Rayleigh-Taylor instability(RTI) (Wolf, 1970; Troyon & Gruber, 1971; Boris, *et al*., 1977; Betti, *et al*., 1993; Piriz, *et al*., 2010; Piriz, *et al*., 2011; Kawata, 2012; Kawata & Karino, 2015): by an additional oscillating acceleration $\delta g$, the RTI growth is mitigated and the RTI perturbation growth is significantly reduced. In this paper, we propose to realize the mitigation mechanism by pulsating and dephasing HIBs in the HIB target implosion. Each HIB has its pulsating phase depending on the HIB axis position in order to produce the global controlled $\delta g$ to mitigate the implosion non-uniformity. Our fluid implosion simulations demonstrate that the implosion acceleration is successfully modulated by the pulsating and dephasing HIBs' illumination, and the controlled $\delta g$ was created during the target implosion.

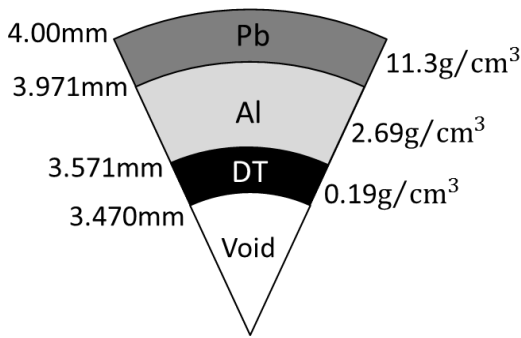

Fig.1 Target structure

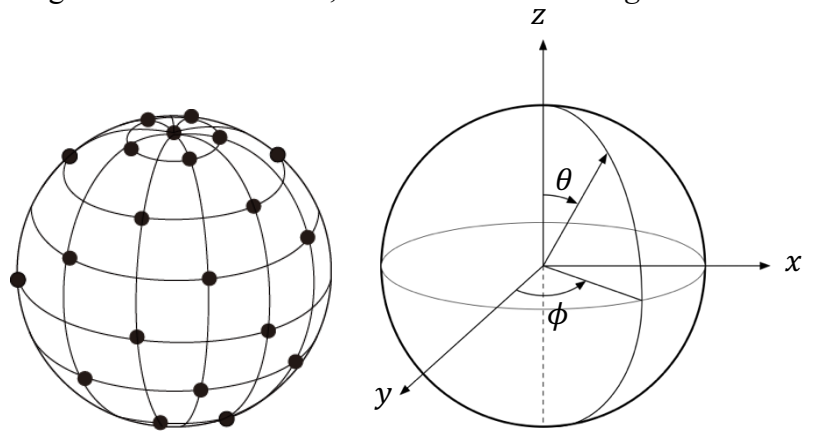

Fig.2 32HIBs system

## Non-uniformity mitigation method

In this study, we analyze the implosion non-uniformity by the arbitrary Lagrangian-Eulerian (ALE) method hydrodynamics simulation (Hirt *et al.*, 1974). The target structure is shown in Fig. 1. The 32 Pb$^+$ ion beams are illuminated in the arrangement shown in Fig.2 to the target (Skupsky & Lee, 1983; Ogoyski *et al.*, 2004; Ogoyski *et al.*, 2010). The HIB particle density has the Gaussian distribution, and the transverse beam emittance is 3.2mm-mrad.

When the instability driver wobbles uniformly in time, the imposed perturbation $\delta g(t)$ for a gravity $g_0$ at $t = \tau$ may be written as

$$g = g_0 + \delta g(t) = g_0 + \delta g e^{i\Omega\tau} e^{\gamma(t-\tau)+i\vec{k}\cdot\vec{x}} \quad (1)$$

Here, $\delta g$ is the amplitude, $\Omega$ the wobbling or oscillation frequency defined actively by the dribing wobbler, and $\Omega\tau$ the phase shift of superimposed perturbations. At each time *t*, the wobbler or the modulated driver provides a new perturbation with the phase and the amplitude actively defined by the driving wobbler itself. The superposition of the perturba- tions provides the actual perturbation at *t* as follows:

$$\int_0^t d\tau\, \delta g e^{i\Omega\tau} e^{\gamma(t-\tau)+i\vec{k}\cdot\vec{x}} \propto \frac{\gamma+i\Omega}{\gamma^2+\Omega^2} \delta g e^{\gamma t} e^{i\vec{k}\cdot\vec{x}} \quad (2)$$

When $\Omega \gg \gamma$, the perturbation amplitude is reduced by the factor of $\gamma/\Omega$, compared with the pure instability growth ($\Omega = 0$) based on the energy deposition non-uniformity. The result in Eq.(2) presents that the perturbation phase should oscillate with $\Omega \gtrsim \gamma$ for the effective amplitude reduction.

In the simulations, we realize the oscillating $\delta g$ and the mitigation mechanism by the following pulsating foot and main HIB pulses. The foot pulse power $P_{foot}$ and the main pulse power $P_{main}$ (see Fig. 3) are represented by the following equations:

$$P_{foot} = 5.80[\text{TW}]\left(1 + A\sin\left(\frac{2\pi t}{T} + \frac{2\pi\xi}{360}\right)\right) \quad (3)$$

$$P_{main} = 320[\text{TW}]\left(1 + A\sin\left(\frac{2\pi t}{T} + \frac{2\pi\xi}{360}\right)\right) \quad (4)$$

Here *A* is amplitude of input pulse, *T* is the pulsation period and $\xi$ is the phase of each pulsating HIB. In this case we employ *A*=0.100 and *T*=1.00ns for our simulations. The phase $\xi$ for each HIB is listed in Table 1.

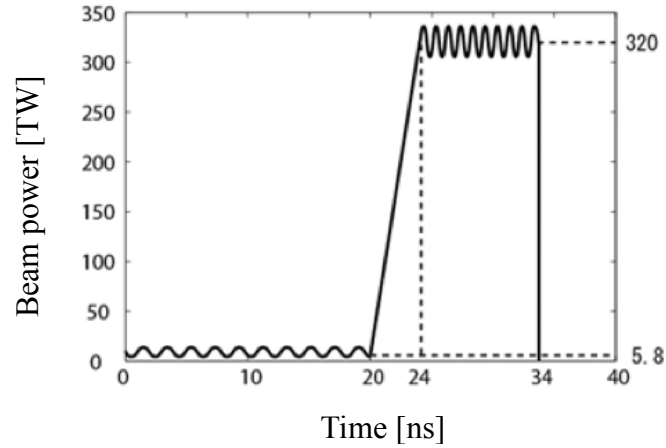

Fig. 3 Beam power pulsation

Table 1 Beam power phase $\xi$ for each HIB

| $\theta$[deg] | $\phi$[deg] | $\xi$[deg] | $\theta$[deg] | $\phi$[deg] | $\xi$[deg] |
|---|---|---|---|---|---|
| 0.000 | 0.000 | 0.000 | 100.812 | 36.000 | 205.714 |
| 37.377 | 0.000 | 51.429 | 100.812 | 108.000 | 277.714 |
| 37.377 | 72.000 | 123.429 | 100.812 | 180.000 | 349.714 |
| 37.377 | 144.000 | 195.429 | 100.812 | 252.000 | 61.714 |
| 37.377 | 216.000 | 267.429 | 100.812 | 324.000 | 133.714 |
| 37.377 | 288.000 | 339.429 | 116.565 | 0.000 | 257.714 |
| 63.435 | 36.000 | 102.857 | 116.565 | 72.000 | 329.143 |
| 63.435 | 108.000 | 174.857 | 116.565 | 144.000 | 41.143 |
| 63.435 | 180.000 | 246.857 | 116.565 | 216.000 | 113.143 |
| 63.435 | 252.000 | 318.857 | 116.565 | 288.000 | 185.143 |
| 63.435 | 324.000 | 30.857 | 142.623 | 36.000 | 308.571 |
| 79.188 | 0.000 | 154.286 | 142.623 | 108.000 | 20.571 |
| 79.188 | 72.000 | 226.286 | 142.623 | 180.000 | 92.571 |
| 79.188 | 144.000 | 298.286 | 142.623 | 252.000 | 164.571 |
| 79.188 | 216.000 | 10.286 | 142.623 | 324.000 | 236.571 |
| 79.188 | 288.000 | 82.286 | 180.000 | 0.000 | 0.000 |

Evaluation method for non-uniformity

In this study, we use the root mean square (RMS) shown by the following equation for the non-uniformity evaluation.

$$\sigma_i^{rms} = \frac{1}{F}\sqrt{\frac{\Sigma_j(F_{ij}-\langle F_{ij}\rangle)^2}{\theta_{mesh}}} \qquad (5)$$

$F_{ij}$ is a physical quantity, $\langle F_{ij}\rangle$ is the average of physical quantity of circumferential direction on a certain radius, $\theta_{mesh}$ is the total mesh number in the $\theta$ direction, and ($i$, $j$) is the mesh numbers for the radial direction and the azimuthal direction, respectively.

We also perform the mode analysis for the non-uniformity $f(\theta)$ based on the Legendre polynomial $P_n$. Here the amplitude of the mode $n$ is obtained by the following equation:

$$A_n = \frac{2n+1}{2}\int_0^\pi f(\cos\theta)P_n(\cos\theta)\sin\theta d\theta \qquad (6)$$

The Legendre polynomial $P_n(P_0 \sim P_5)$ is shown in Fig.4 for reference.

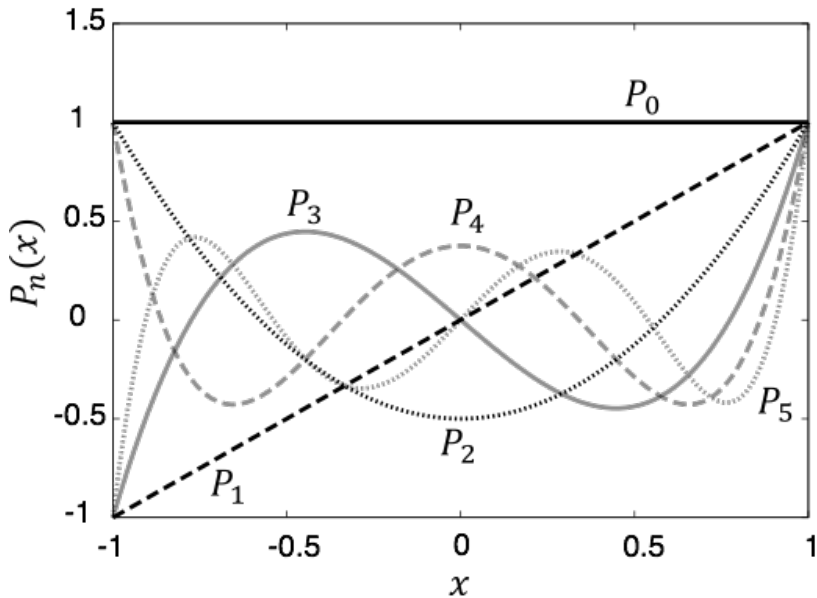

Fig.4 Legendre polynomial $P_n(P_0 \sim P_5)$

Non-uniformity mitigation in HIF target implosion

First, we examine the effect of the pulsating and dephasing HIBs illumination on the target implosion acceleration. Figures 5 show the implosion acceleration histories at $\theta = 0 \text{ deg}$ and $152 \text{ deg}$ for the DT layer. In Fig. 5(a) the pulsating HIBs are in phase, and so the implosion acceleration is also in phase. However, in Fig. 5(b) the pulsating HIBs' phases are controlled as shown in Table 1. Figure 5(b) demonstrates that the pulsating and controlled dephasing HIBs illumination creates the DT fuel imlosion acceleration oscillation of $\delta g$.

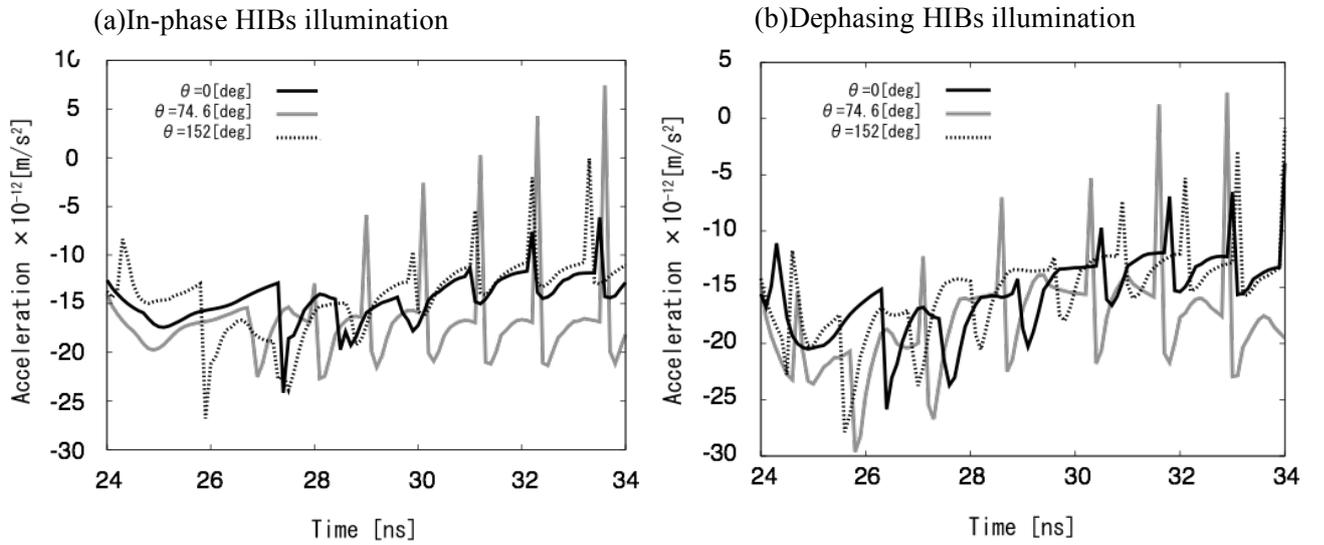

Fig. 5 Time histories of the DT fuel acceleration.

Table 2 shows the summary of the implosion simulation results for the in-phase HIBs illumination and for the dephasing HIBs. The results with vibration beam has short Void close time, more Gain and Max$\rho$ but its Max$T_i$ was decreaced.

Table 2 Implosion result summary for the in-phase HIBs and for the dephasing HIBs

|  | In-phase HIBs | Dephasing HIBs ($A = 0.100$) |
|---|---|---|
| Void close time [ns] | 38.9 | 38.2 |
| Gain | 38.7 | 50.8 |
| Max $\rho$ [kg/m$^3$] | 21800 | 21900 |
| Max $T_i$ [KeV] | 9.23 | 8.48 |

Figures 6 show the non-uniformity histories of the density $\rho$, the ion tempareture $T_i$, the pressure $P$ and the radial direction speed $V_r$ at the DT layer. The solid line shows the non-uniformities by the in-phase HIBs' illumination, and the dotted line shows them by the dephasing pulsating HIBs' illumination. Figures 6 present that the dephasing and pulsating HIBs reduce the non-uniformities successfully.

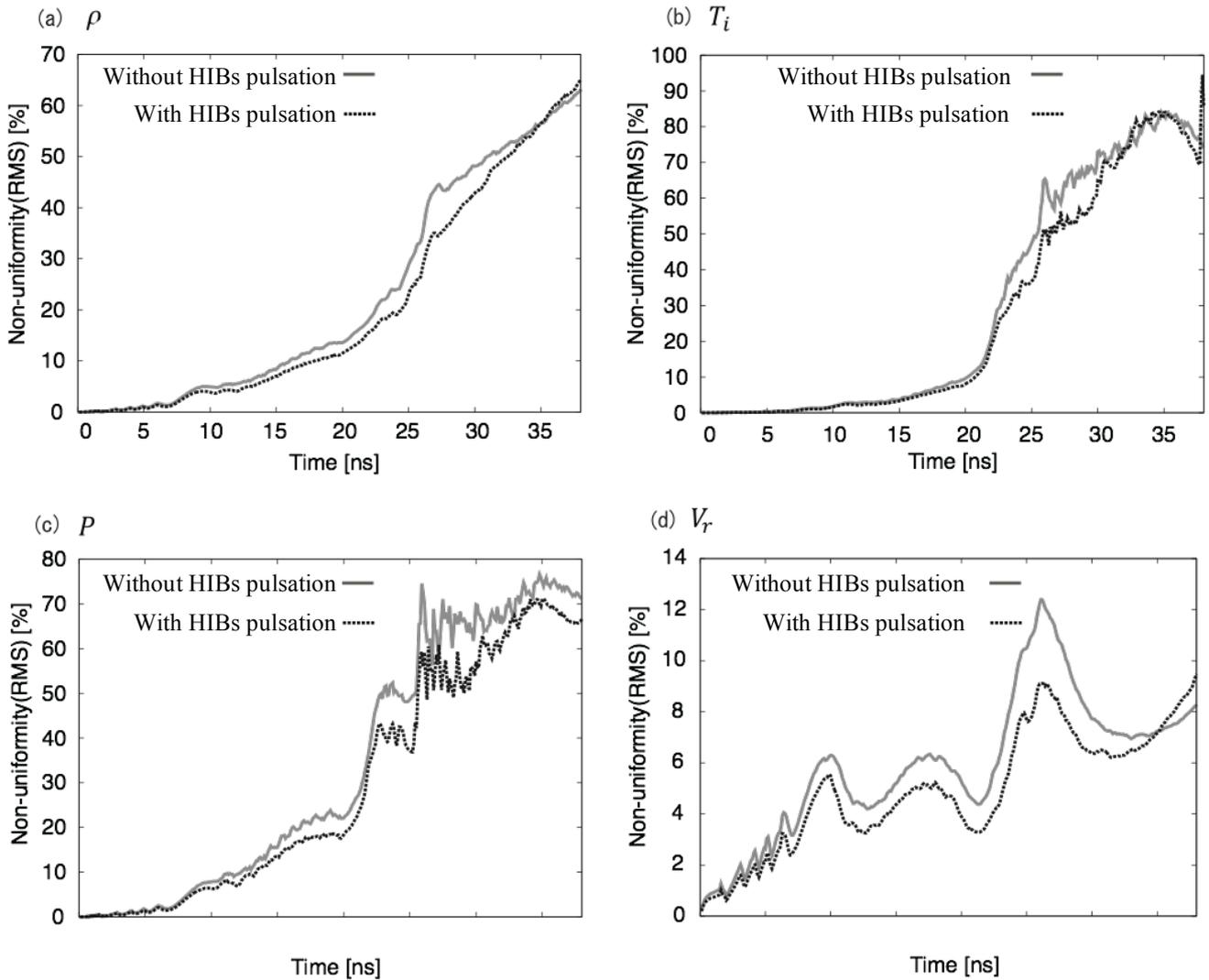

Fig. 6 Time histories of the DT fuel non-uniformities for the pulsating and dephasing HIBs and for the HIBs without the pulsation.

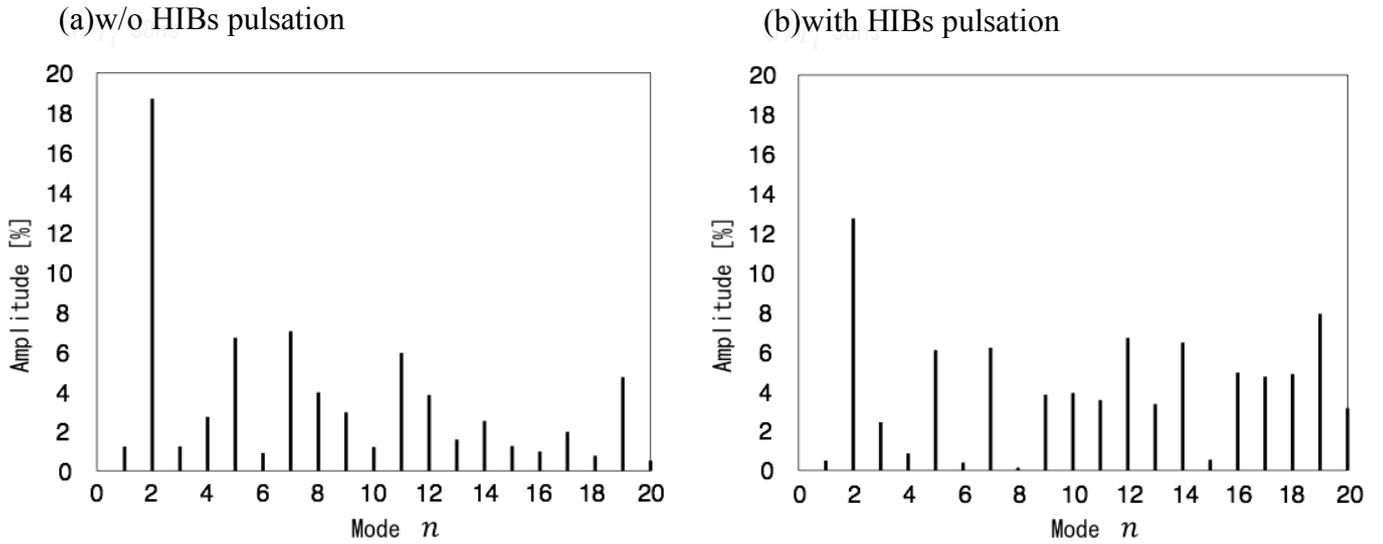

(a) w/o HIBs pulsation  (b) with HIBs pulsation

Fig. 7 Modes of the ion temperature $T_i$ in the DT layer at t=38ns.

Figures 7(a) and (b) show the non-uniformity mode analyses results for the averaged ion temperature $T_i$ of the DT layer at $t = 38$ns. Figures 7 present that the dephasing and pulsating HIBs reduce the largest mode of the "Mode 2" significantly.

In Table 1, Figs. 5, 6 and 7(b), the oscillation amplitude $A$ of the HIBs input power in Eqs. (2) and (3) was $A$=0.100. Figure 8 shows the relation of the fuel target gain versus the oscillation amplitude $A$ of the HIBs input power. When the HIBs input power oscillation of $A$ is 0.1, the fuel target gain becomes the maximum. The target gain becomes 0, when $A$ exceeds 0.140. The results in Fig. 8 demonstrates that the dephasing and pulsating HIBs illumination realizes the better uniformity in the DT fuel implosion, and consequently leads a higher gain.

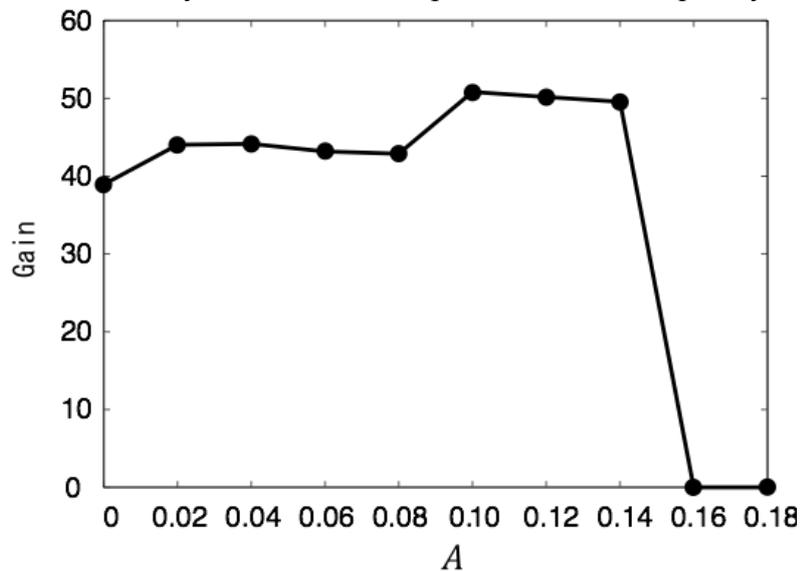

Fig. 8 Target fuel gain versus the HIBs pulsating amplitude $A$.

We also study the robustness against the displacement *dz* (see Fig.9) of the target misalignment in a fusion reactor. Figure 10 shows the relation between the fuel target gain and the displacement *dz* for each input pulse modulation amplitude *A*. Figure 10 presents that the dephasing and pulsating HIBs illumination is robust against the target misalignment *dz*.

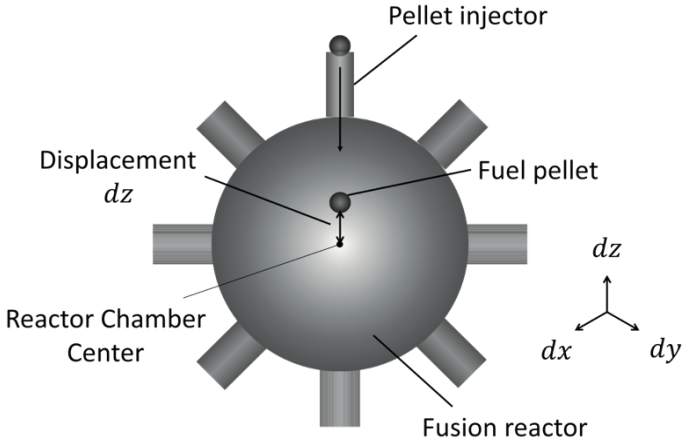 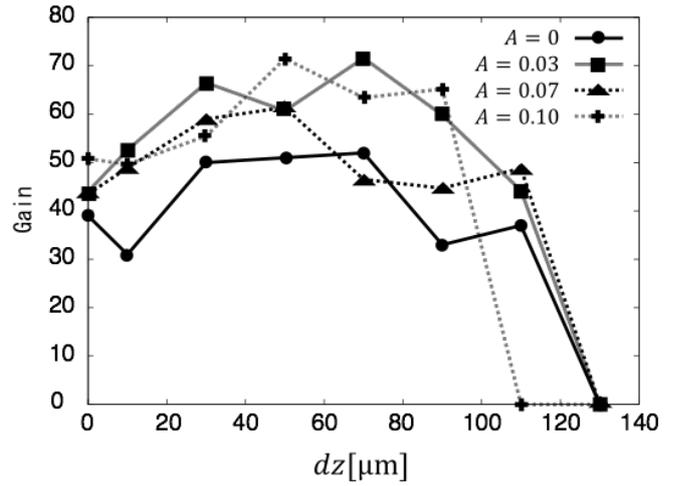

Fig.9 Target displacement *dz*   Fig. 10 Gain versus the target misalignment *dz*.

Conclusions

In this paper we have shown that the pulsating and dephasing HIBs illumination creates the oscillating acceleration $\delta g$, which mitigates the RTI growth. In our previous works (Kawata, *et al.*, 1993; Kawata, 2012,; Kawata & Karino 2015) it was demonstrated that the oscillating acceleration $\delta g$ reduces the instability growth significantly. The pulsating HIBs illumination onto a fuel target induces the oscillating $\delta g$ successfully. It was found that the target material responds to the deposited HIBs pulsation directly. In this paper the pulsating HIBs phases are desinged as shown in Table 1 to create the large wave mode of $P_2$ or so, so that the RTI growth rate would be also minimized. The work presented in this paper demonstrates that the controlled HIBs illumination provides a useful tool to realize a stable and uniform implosion in HIF.